# FBG-Based Triaxial Force Sensor Integrated with an Eccentrically Configured Imaging Probe for Endoluminal Optical Biopsy

Zicong Wu, Anzhu Gao, *Member, IEEE*, Ning Liu, *Student Member, IEEE*, Zhu Jin and Guang-Zhong Yang, *Fellow, IEEE*

*Abstract*— Accurate force sensing is important for endoluminal intervention in terms of both safety and lesion targeting. This paper develops an FBG-based force sensor for robotic bronchoscopy by configuring three FBG sensors at the lateral side of a conical substrate. It allows a large and eccentric inner lumen for the interventional instrument, enabling a flexible imaging probe inside to perform optical biopsy. The force sensor is embodied with a laser-profiled continuum robot and thermo drift is fully compensated by three temperature sensors integrated on the circumference surface of the sensor substrate. Different decoupling approaches are investigated, and nonlinear decoupling is adopted based on the cross-validation SVM and a Gaussian kernel function, achieving an accuracy of 10.58 mN, 14.57 mN and 26.32 mN along X, Y and Z axis, respectively. The tissue test is also investigated to further demonstrate the feasibility of the developed triaxial force sensor.

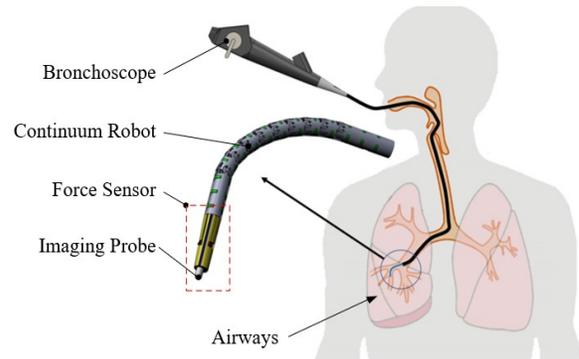

Figure 1. The developed force sensor is integrated with the continuum robot for the optical biopsy toward the distal airways to diagnose the early-stage lung cancer [6].

## I. INTRODUCTION

Endoluminal intervention is an increasingly popular approach to minimally invasive diagnosis and treatment of early-stage cancers [1]. To this end, continuum robots represent an ideal choice for these tasks [2]. Various surgical instruments, like needles or forceps, have also been developed so that they can be inserted through the inner lumen of the continuum robot to perform desired surgical operations. Recently, the therapeutic approaches are increasingly shifted towards the use of optical biopsies to identify potential precancerous tissue [3]. Compared to traditional re-sectional biopsy, the optical biopsy techniques enable in situ, in vivo diagnosis and treatment in a single step [4]. The probe-based confocal laser endomicroscopy (pCLE) is a promising in-vivo imaging technique for this purpose [5].

Recently, we have developed a laser-profiled continuum robot to steer a flexible imaging probe to the distal airways for the diagnosis of early-stage lung cancer at distal bronchi [6, 7], as shown in Figure 1. During optical biopsy, the contact force between the imaging probe and the tissue takes an important role to guarantee high quality image acquisition, whilst maintaining safe interaction with the anatomy. The importance of force-sensing during endoluminal interventions has been studied and verified in [8, 9]. It has been further illustrated in [10, 11] that a contact force ranging from $0.1\,N$ to $0.5\,N$ enables a stable optical scanning for images acquisition using pCLE. However, the complex endoluminal environment poses additional challenges to traditional approaches to force sensing. Similar work such as [12] focuses on the development of contact force sensor using the carbon-nanotube-coated 3D micro-spring, but its current state is still not ready for being applied for pCLE tissue scanning task. In recent years, force sensing using FBG (Fibre Bragg Gratings) sensors become an ideal option for continuum robots, due to its advantages including high sensitivity, high accuracy, biocompatibility, electrical passivity, chemical inertness, as well as its miniature size [13]. This provides us a potential solution to overcome such issues.

Thus far, there are some existing force sensors based on the FBG sensors to accomplish triaxial force sensing with thermo drift compensation for medical applications. A force sensor with three lateral FBG sensors and a central FBG sensor was developed in [14 -16] to accomplish the sensing of triaxial force exerted between the surgical tools and retinal tissues for the robotic-assisted retinal surgery. Our previous work in [16] designed a force sensor, using parallel flexure hinges and four FBG sensors configured inside to achieve triaxial force sensing with balanced resolution along each axis. However, those sensors have to take the inner space or lumen to place another FBG sensor for the thermo drift compensation, which is difficult to be combined with our sensing demand of optical biopsy. Our aim is to configure the imaging probe eccentrically inside the continuum robot as well as the sensor substrate, and all the sensing components should only be

This work was supported by Engineering and Physical Sciences Research Council (EPSRC), United Kingdom (EP/N019318/1). Anzhu Gao is also supported by SJTU Global Strategic Partnership Fund (2019 SJTU-CUHK) and the State Key Laboratory of Robotics and Systems (HIT) (SKLRS-2020-KF-18). (*Corresponding authors: Anzhu Gao and Guang-Zhong Yang*)

Z. Wu, N. Liu, Z. Jin are with the Hamlyn Centre for Robotic Surgery, Imperial College London, SW7 2AZ, London, UK.

A. Gao is with the Institute of Medical Robotics and Department of Automation, Shanghai Jiao Tong University, and the Key Laboratory of System Control and Information Processing, Ministry of Education, Shanghai 200240, China.

G. Z. Yang is with the Institute of Medical Robotics, Shanghai Jiao Tong University, 200240, Shanghai, P. R. China.

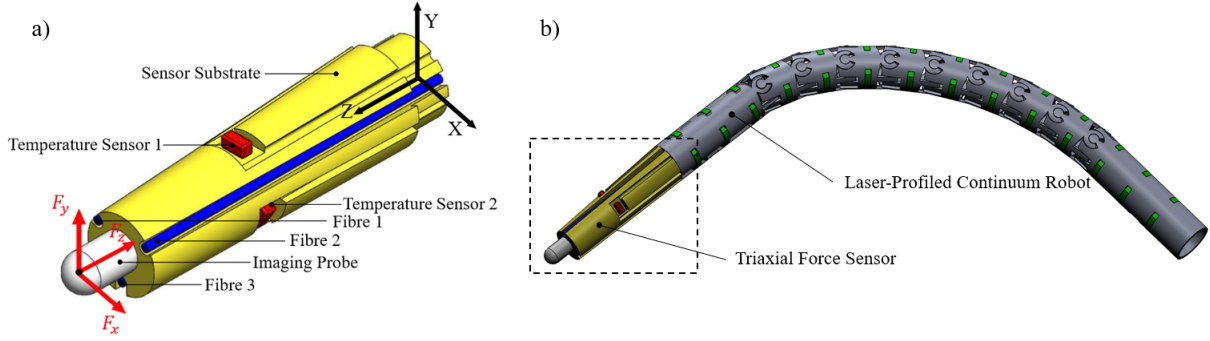

Figure 2. The structure of the developed force sensor and the laser-profiled continuum robot. a) displays the detailed structure and configuration of the force sensor with thermo drift compensation. It consists of a tapered substrate with an eccentrically placed imaging probe, three FBG sensors and three temperature sensors that are evenly distributed on the circumference surface of the substrate; b) shows the integration of force sensor and the laser-profiled continuum robot.

placed on the outer surface without occupying the inner space. This poses a challenge to the design of a triaxial force sensor integrated with our miniaturised laser-profiled continuum robot.

In this paper, a triaxial force sensor is developed by using a tapered substrate with an eccentric inner lumen to hold the imaging probe inside and sensing elements outside, whilst being compatible with the continuum robot. Various decoupling approaches including linear and nonlinear, regression-based and learning-based methods are adopted to achieve accurate force sensing with thermo drift compensation successfully. This paper is organized as follows. Section II illustrates the design principle and theoretical calculation; Section III presents the experimental setup, data acquisition, and calibration methods; Section IV demonstrates the evaluation results and the effectiveness of the method.

## II. DESIGN AND THEORETICAL CALCULATION

This section presents the design of the sensor substrate, which is key to sensor integration and providing sensitive strain measurements with the desired resolution. The associated theoretical calculation is provided in detail to verify the effectiveness of the proposed method.

### A. Substrate Design

Traditionally, triaxial force sensing with thermo drift compensation is achieved by using four fibres, with three distributed on the circumference surface of the sensor substrate, and one located at the central lumen. However, to perform optical biopsy based on our developed continuum robot, the inner lumen has been occupied by an imaging probe that is configured eccentrically. To accomplish triaxial force sensing on the tip of the imaging probe and guarantee the compatibility between them and the continuum robot, a truncated oblique cone-shaped substrate has been designed. Three miniature temperature sensors are placed evenly on the circumference surface of the substrate for thermo drift compensation in real time. This substrate with a thin-walled tube structure is fabricated by 3D printing using the *formlab* printer, as shown in Figure 2.

The total length $L$ of the sensor substrate is 8 $mm$ with a 1 $mm$ shaft to facilitate the assembly with the continuum robot. A lumen with a diameter of $d$ and three channels with a diameter of $2r$ are used to place the imaging probe and three fibres separately. Three temperature sensors are distributed evenly on the surface wall, located at the same axial position with FBG sensors to obtain endoluminal temperature for compensation. The distance $\Delta L$ from them to the top end along Z-axis is 3 $mm$. The diameter of the top end $D_{top}$ and bottom end $D_{bot}$ are 1.8 $mm$ and 2.2 $mm$, respectively. Dimensions of the grooves and channels to place temperature sensors and wires are negligible to simplify the following theoretical calculation.

### B. Theoretical Calculation

Based on the Euler-Bernoulli beam theorem, the strain of each fibre with three-dimensional load $F_x$, $F_y$, and $F_z$ exerting on the tip of the imaging probe can be derived as $\epsilon_1$, $\epsilon_2$ and $\epsilon_3$, in the following equations:

$$\epsilon_1 = \frac{(\Delta L + y_F) \cdot R \cdot \sin\beta}{E \cdot I_x} \cdot F_x + \frac{(\Delta L + y_F) \cdot (R \cdot \cos\beta + \Delta y)}{E \cdot I_y} \cdot F_y - \frac{1}{E \cdot A} \cdot F_z \quad (1)$$

$$\epsilon_2 = -\frac{(\Delta L + y_F) \cdot R \cdot \sin\alpha}{E \cdot I_x} \cdot F_x + \frac{(\Delta L + y_F) \cdot (R \cdot \cos\alpha + \Delta y)}{E \cdot I_y} \cdot F_y - \frac{1}{E \cdot A} \cdot F_z \quad (2)$$

$$\epsilon_3 = \frac{(\Delta L + y_F) \cdot (R - \Delta y)}{E \cdot I_y} \cdot F_y - \frac{1}{E \cdot A} \cdot F_z \quad (3)$$

where:

$$I_x = \frac{\pi \cdot D^4}{64} + \frac{\pi \cdot D^2 \cdot y_c^2}{4} - \frac{\pi \cdot d^4}{64} - \frac{\pi \cdot d^2 \cdot \Delta y^2}{4} - \frac{3 \cdot \pi \cdot r^4}{4} \quad (4)$$
$$- \pi \cdot r^2 \cdot ((R \cdot \cos\alpha + \Delta y)^2 + (R \cdot \cos\beta + \Delta y)^2 + (R - \Delta y)^2)$$

$$I_y = \frac{\pi \cdot D^4}{64} - \frac{\pi \cdot d^4}{64} - \frac{3 \cdot \pi \cdot r^4}{4} - \pi \cdot r^2 \cdot ((R \cdot \sin\alpha)^2 + (R \cdot \sin\beta)^2) \quad (5)$$

$$A = \frac{\pi}{4} \cdot (D^2 - d^2) - 3 \cdot \pi \cdot r^2 \quad (6)$$

$$D = D_{top} + \frac{\Delta L}{L} \cdot (D_{bot} - D_{top}) \quad (7)$$

$$y_c = \frac{d^2 \cdot \Delta R}{D^2 - d^2} \quad (8)$$

$$\Delta y = \frac{d^2 \cdot \Delta R}{D^2 - d^2} + \Delta R \quad (9)$$

In the above equations, $\Delta R$ is the distance between the centre points of the top end and the central lumen, $R$ is the distance between channels for placing fibres and the centre point of the top-end, $y_c$ is the coordinate in Y-axis of the top-end's centroid, $y_F$ is the distance along Z-axis between the imaging probe tip and the location of the FBG sensor. α is the angel between Fibre 1 and the Y- axis while β is the angle between Fibre 2 and the Y-axis, respectively.

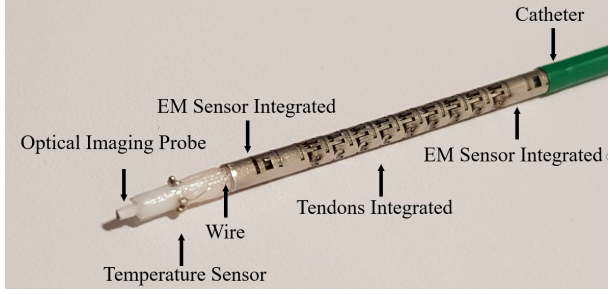

Figure 3. The prototype of the fabricated force sensor assembled with our developed laser-profiled continuum robot.

Figure 2 a) shows the design of the force sensor with a tapered substrate, three FBG sensors and three temperature sensors, and an imaging probe eccentrically configured inside. According to the properties of the FBG sensors, the wavelength detected is decided by both strain and temperature shift as (10):

$$\Delta\lambda/\lambda = K_\epsilon \cdot \epsilon + K_T \cdot \Delta T \qquad (10)$$

where $\lambda$ is the wavelength of the incident light, $\Delta\lambda$ is the detected wavelength shift, $K_\epsilon$ and $K_T$ are the parameters associated with strain and temperature sensitivities respectively, $\epsilon$ and $\Delta T$ are the strain and temperature shift separately. Therefore, a theoretical model to predict the force exerted at the probe tip can be derived. The force components along each axis can be revealed by the detected wavelength shift of the incident light in each fibre by the interrogator and obtained temperature change as (11):

$$\begin{bmatrix} F_x \\ F_y \\ F_z \end{bmatrix} = M \cdot [\Delta\lambda_1 \quad \Delta\lambda_2 \quad \Delta\lambda_3 \quad \Delta T_1 \quad \Delta T_2 \quad \Delta T_3]^T \qquad (11)$$

where $M$ is the mapping matrix, $\Delta\lambda_1$, $\Delta\lambda_2$, $\Delta\lambda_3$ are the wavelength shift in each fibre, $\lambda_1$, $\lambda_2$, $\lambda_3$ are the wavelength of incident light in each fibre, and $\Delta T_1$, $\Delta T_2$, $\Delta T_3$ are the temperature shift derived from the readings of three temperature sensors. The units of force components, wavelength shift, and temperature change are $N$, $nm$ and $K$, respectively.

## III. EXPERIMENTAL SETUP, DATA ACQUISITION AND CALIBRATION

This section introduces the experimental platform and the steps involved for data transmission, collection and processing in order to obtain a calibrated model.

### A. Experimental Setup

Three optical fibres with Ormocer coating (190 $\mu m$, single 3 $mm$ DTG, FBGS International) and three miniature temperature sensors (ERTJZEG103FA Thermistor, Panasonic Corporation) are integrated on the substrate by glue. Then the fabricated sensor is assembled with our developed laser-profiled continuum robot [17], as shown in Figure 3. This unit is fixed at the end of a catheter to encapsulate all fibres, tendons and wires to guarantee the safety of these fragile components and then placed on the experimental platform, as shown in Figure. 4. The force sensor can be actuated by the lead screw to contact the specimen placed on the weighing surface of the electronic scale, making the force loaded on the force sensor. A series of location holes on the platform enables the adjustment of the sensor's orientation and position thus varying force components along each axis can be loaded on the sensor. The fibres are connected to the interrogator (FBG-Scan 804D, FBGS International) to detect the wavelength shift of each FBG sensor under the applied load. In addition, the readings of temperature sensors are measured by the DAQ device (NI USB X series 6356, National Instruments) and the applied force is measured by the electronic scale (WTC 600, RADWAG Balances and Scales) with a resolution of 0.005g.

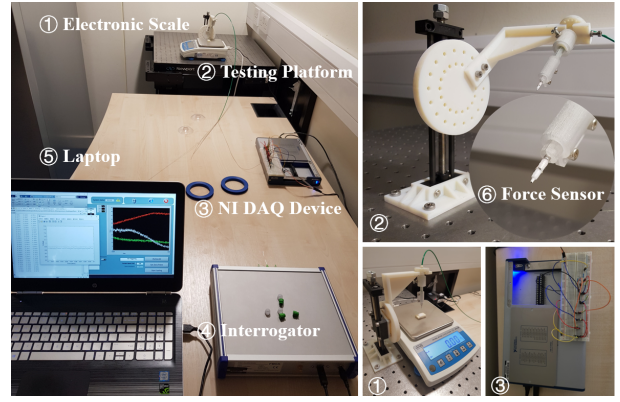

Figure 4. Experimental setup for data acquisition during the processes of both calibration and validation. It includes a high-resolution electronic scale, a NI DAQ device, an FBG interrogator and a 3D printed testing platform for force loading experiments.

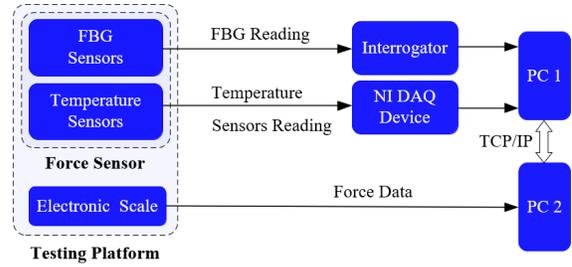

Figure 5. The architecture of data transmission between devices. The readings from the force sensor are measured by the interrogator and NI DAQ device and then collected by PC1, with the force recorded by PC2.

### B. Data Acquisition

The architecture of data transmission between these device is illustrated in Figure 5. The interrogator and the DAQ device are both connected to PC 1 for raw data acquisition. Wireless transmission between PC 1 and PC 2 is achieved by the TCP/IP communication. The datasets are packaged and transmitted to PC 2 to determine and visualize sensed force. The electronic scale is connected to PC 2 directly to collect force data. After deriving force components along each axis based on the position information, real-time contrast can be provided to validate the force sensing of calibrated models.

### C. Calibration

The calibration process is to generate an accurate model to describe the sensor's performance using the collected data. This part consists of the calibration of temperature sensors and approaches to force sensor calibration.

TABLE I. THE CALIBRATION RESULTS OF TEMPERATURE SENSORS

| No. | Equation | R-Square | RMSE |
|---|---|---|---|
| Sensor1 | $\ln(U) = 1545 \times 1/T - 3.08$ | 0.96 | 0.03 |
| Sensor2 | $\ln(U) = 1510 \times 1/T - 2.93$ | 0.97 | 0.02 |
| Sensor3 | $\ln(U) = 1570 \times 1/T - 3.15$ | 0.95 | 0.03 |

TABLE II. THE STATISTICAL ANALYSIS ABOUT FORCE SENSING BASED ON LINEAR CALIBRATED MODELS

| Axis | Resolution/mN | Accuracy/mN | STD of Residual Errors/mN |
|---|---|---|---|
| X | 0.47 | 18.97 | 18.92 |
| Y | 0.38 | 27.41 | 24.51 |
| Z | 1.53 | 57.26 | 45.21 |

*1) Calibration of Temperature Sensors*

Based on theoretical calculation, wavelength shift and temperature change can be linearly mapped to the triaxial force components. However, due to its nonlinear performance, the readings from the temperature sensors can't be utilized directly for linear calibration. Thus, calibration of the temperature sensors is required to derive the relationship between the readings and actual temperature being measured. To this end, the temperature sensors are well-encapsulated and placed into hot water to obtain different environment temperature. Then the temperature is measured by a thermal reader when the reading maintains constant. Therefore, the readings of each temperature sensor as well as a series of corresponding temperatures are obtained. Referring to the datasheet, the relationship between the temperature and the reading can be described by equation (12):

$$R = R_0 \cdot e^{(B \cdot (\frac{1}{T} - \frac{1}{T_0}))} \tag{12}$$

where $T_0$ is the normal temperature of 298.15 $K$, $R_0$ is the resistance at $T_0$, $B$ is the thermal constant ($K$). By recording the readings of temperature sensors in different temperatures, the results are shown in TABLE I, where $U$ within each equation is the raw readings from temperature sensors.

*2) Linear Calibration*

After calibrating these temperature sensors, readings from the DAQ device can be used to derive the mapping matrix. An initial position is set to facilitate acquiring the force sensor's position information. Afterwards, force components along each axis can be derived. Based on (10), when there is no wavelength and temperature shift, the applied force would be zero. Thus, before applying the load, the sensor should maintain a stable state without reading shift, afterwards, the initial reading is used to calculate wavelength shift and temperature change. Although in practice, fluctuation always exists and can't be eliminated, the effect on the sensor's performance is negligible. All datasets are reorganized as the input matrix and output matrix that is consisted of sensor readings and measured force components, separately. They are mapped by a mapping matrix $M$ derived by linear regression:

$$M = \begin{bmatrix} -0.182 & -0.117 & 0.476 & -0.020 & -0.020 & 0.044 \\ -0.313 & 0.381 & 0.062 & 0.008 & -0.014 & 0.001 \\ -1.241 & -1.443 & -1.538 & 0.116 & 0.096 & 0.081 \end{bmatrix} \tag{13}$$

*3) Nonlinear Calibration*

Theoretically, linear equations can be derived to describe the relationship between triaxial force components and detected wavelength and temperature readings. However, in practice, many factors can lead to nonlinearities. For example, the nonuniform gluing between the fibres and the substrate, manufacturing error on the substrate and inadequate assembly. Thus, some nonlinear models have been developed to introduce nonlinearities into the model in different degrees to improve the performance of force sensing. The regression-based nonlinear calibration introduces the contribution of intercept terms, quadratic terms of each predictor, and the joint production of any two predictors. Different models have been used to adjust datasets from sensor readings for better fitting, including linear, pure quadratic and quadratic models. Inspired by [18, 19], the learning-based calibration method is developed, the SVM (Support Vector Machine) algorithm has been implemented to learn from training datasets using various kernel functions, including linear, Gaussian and polynomial.

## IV. RESULTS

The datasets are collected when performing multiple loading and releasing processes, by rotating the lead screw. Then it is divided into training dataset and testing dataset and contains high to 16122 and 4242 sets of data respectively, which is for better regression result and avoiding overfitting when using learning-based calibration methods. This section displays calibrated models derived by multiple approaches and the residual error between the predicted and measured force components along each axis is calculated as accuracy. The effectiveness of force sensing and thermo drift compensation is validated in experiments using these models.

### A. Linear Model

Based on the mapping matrix and theoretical analysis of the force sensor, the resolution along each axis can be derived, with a value of 0.476 $mN$, 0.381 $mN$ and 1.538 $mN$ respectively, along the X, Y and Z axes. From TABLE II, it can be observed that the accuracy of force sensing along the Z axis is not as good as that along X and Y axes. This indicates a relatively poor resolution along Z axial, i.e., for the same wavelength shift, the deviation is larger, which is due to the structure and the material properties of the substrate.

### B. Nonlinear Model

To further improve the performance of force sensing, both traditional regression-based and learning-based calibration methods have been applied. For the regression-based calibration method, the training data is re-adjusted to generate new training datasets for better regression. Several adjusting models have been applied for optimizing the calibration result. By using the linear model to adjust datasets, the calibration mode is derived as (14), (15), (16), with intercept terms introduced. When utilizing the pure-quadratic model, except for the intercept terms, the quadratic terms of each predictor is also involved, with results displayed as (17), (18) and (19). Also, the quadratic model introduces up to 28 terms into the model including intercept terms, quadratic terms and joint production of any two predictors, as (20), (21) and (22) shows:

$$F_x = 0.500 \cdot \sum_i^7 \text{Coefficient}_i \cdot \text{Term}_i \tag{14}$$

$$F_y = 0.492 \cdot \sum_i^7 \text{Coefficient}_i \cdot \text{Term}_i \tag{15}$$

$$F_z = 2.332 \cdot \sum_i^7 \text{Coefficient}_i \cdot \text{Term}_i \tag{16}$$

$$F_x = 0.589 \cdot \sum_i^{13} \text{Coefficient}_i \cdot \text{Term}_i \tag{17}$$

$$F_y = 0.562 \cdot \sum_i^{13} \text{Coefficient}_i \cdot \text{Term}_i \tag{18}$$

TABLE III. THE EFFECT ANALYSIS ABOUT REGRESSION-BASED NONLINEAR CALIBRATION USING VARIOUS ADJUSTING MODELS.

| Adjusting Model | | R-squared | Accuracy/mN | STD of Residual Errors/mN |
|---|---|---|---|---|
| Linear | X | 0.94 | 17.92 | 18.08 |
| | Y | 0.94 | 27.65 | 24.37 |
| | Z | 0.75 | 57.31 | 44.07 |
| Pure Quadratic | X | 0.95 | 17.16 | 15.36 |
| | Y | 0.96 | 24.09 | 19.86 |
| | Z | 0.76 | 55.25 | 42.97 |
| Quadratic | X | 0.98 | **11.19** | **9.62** |
| | Y | 0.99 | **14.18** | **13.18** |
| | Z | 0.84 | **43.09** | **33.38** |

TABLE IV. THE EFFECT ANALYSIS ABOUT LEARNING-BASED NONLINEAR CALIBRATION USING VARIOUS KERNEL FUNCTIONS.

| Kernel Function | | Accuracy/mN | STD of Residual Errors/mN |
|---|---|---|---|
| Linear | X | 17.28 | 17.84 |
| | Y | 26.11 | 26.25 |
| | Z | 57.28 | 46.02 |
| Gaussian | X | **10.58** | **11.33** |
| | Y | **14.57** | **18.64** |
| | Z | **26.32** | **25.73** |

$$F_z = 2.357 \cdot \sum_i^{13} \text{Coefficient}_i \cdot \text{Term}_i \quad (19)$$

$$F_x = 0.658 \cdot \sum_i^{28} \text{Coefficient}_i \cdot \text{Term}_i \quad (20)$$

$$F_y = 1.135 \cdot \sum_i^{28} \text{Coefficient}_i \cdot \text{Term}_i \quad (21)$$

$$F_z = 12.507 \cdot \sum_i^{28} \text{Coefficient}_i \cdot \text{Term}_i \quad (22)$$

The detailed terms and their corresponding coefficients are provided in the Appendix[1] as TABLE V, TABLE VI, and TABLE VII. TABLE III displays the statistical analysis of each derived model. It can be observed that the model derived based on the quadratic adjusting model offers the best accuracy and the lowest standard deviation of residual errors, which means more accurate and robust force sensing. This suggests that the increasing number of terms introduced higher non-linearities into the model, leading to better prediction results.

For the learning-based calibration method using the SVM algorithm, training with different kernel functions are applied, including linear, Gaussian and polynomial. Cross validation is implemented during the training process to avoid overfitting. TABLE IV offers detailed analysis of the force sensing effect of the trained model using different kernel functions.

### C. Validation of Force Sensing

The above sections validate the effect of various models derived by different approaches and it can be observed that the best triaxial force sensing is achieved by the model training using the SVM algorithm and the Gaussian kernel function. Thus, this model is applied to provide real-time triaxial force sensing as well as visualize the predicted force components along each axis based on the readings from the force sensor.

To validate triaxial force sensing, a testing environment is built by using thin chicken breast tissue to simulate the anatomy of lung branching system, as shown in Figure 6 a).

[1] https://www.dropbox.com/sh/4rrtl6k3rigzkom/AACDVEANJ4nhX95uwRjQHpbUa?dl=0.

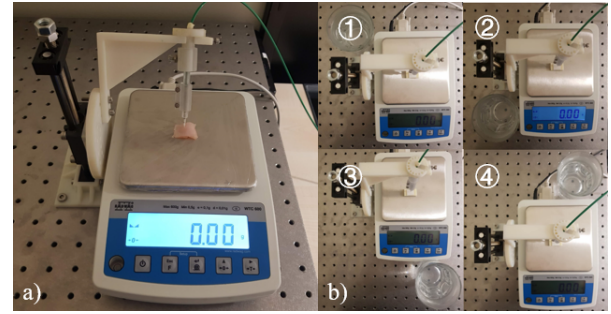

Figure 6. The experimental setup for validating the force sensing and thermo drift compensation. In a), a piece of chicken is placed at the tip of the sensor on the electronic scale to simulate the tissue and in b), a cup of boiled water is placed at position 1,2,3,4 to provide variable thermal environment for the force sensor.

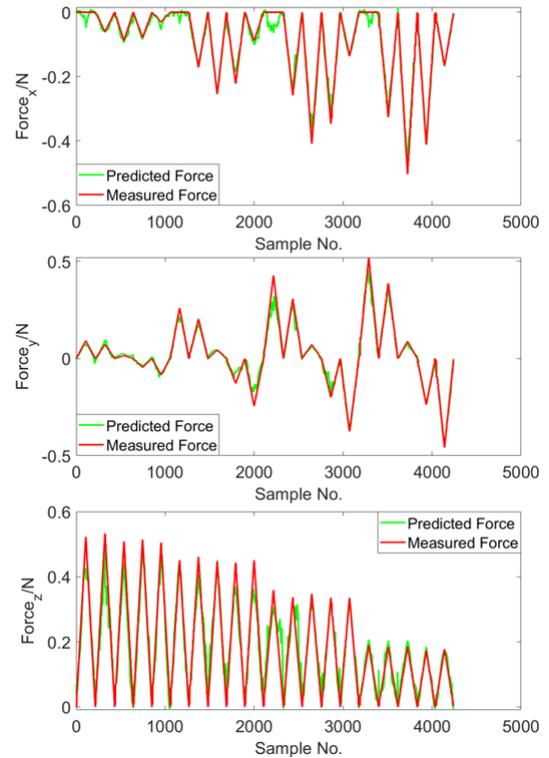

Figure 7. The experimental results for validating force sensing. The predicted and measured triaxial force components are plotted in green and red separately.

The force sensor is actuated to approach and interact with the tissue placed on the weighing surface of the electronic scale to collect force data, with various positions and orientations. In the meantime, the sensing result is determined and visualized in PC 2 as Figure 7. Both the measurement data and sensing data are recorded for further analysis, as shown in Figure 8.

### D. Validation of Thermo Drift Compensation

Generally, the force sensor is assembled at the tip of continuum manipulator, which is intended to use inside the human airways. The temperature inside the human will not have an extreme change due to the controllable and confined operation room during the interventional surgery. Hence, to validate the effect of thermo drift compensation, a similar experimental process is performed with a cup of boiled water placed at different positions of the platform as heat source to change the thermal properties of environment, as shown in Figure 6 b). Because the FBG sensor is extremely sensitive to environment temperature, a nearby heat source would significantly affect its performance. While in the control group no hot water placed and the sensor just work under the environment temperature around 25 °C. The statistical analysis of triaxial residual errors is shown in Figure 9. It indicates that in different thermal environments, the shift of mean and standard deviation of the residual errors along each axis is inconspicuous, demonstrating the excellent effect of thermo drift compensation.

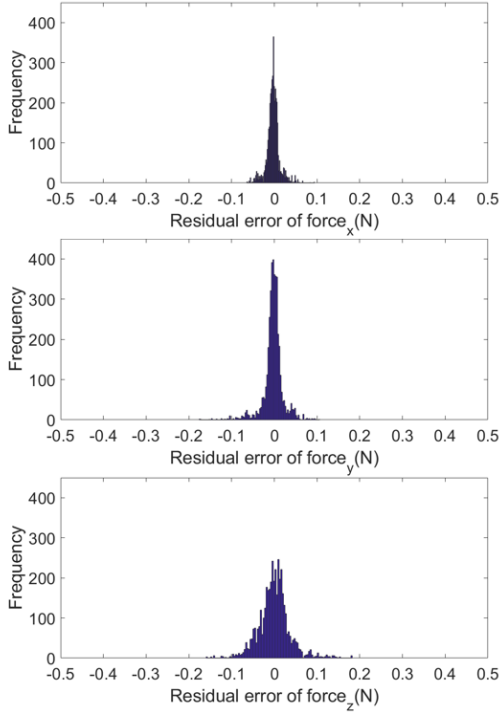

Figure 8. The histogram of residual errors from triaxial force components. From top to bottom, the figures display the distribution of residual errors in X, Y and Z axis respectively.

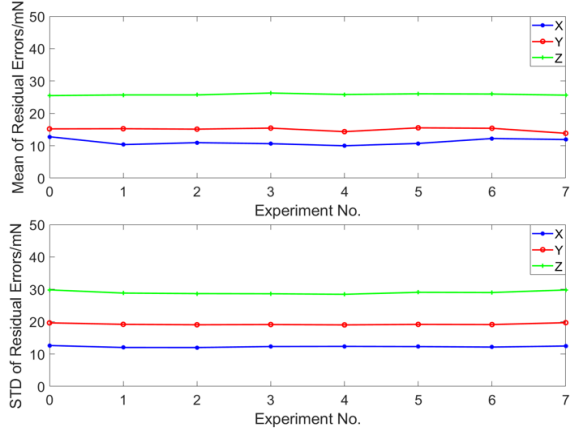

Figure 9. Validation results of thermo drift compensation. The mean and standard deviation of triaxial residual errors from each experiment is plotted respectively. The data of group 0 is from the control group.

## V. Discussion

We have developed a triaxial force sensor for the optical biopsy. The calibration models using different approaches have been successfully derived and validated by residual error analysis. The accuracy along each axis has met our design requirements, of which the accuracy along the Z axis is worse, attributed to the substrate structure. This is also indicated by the resolution analysis in linear calibration, and may be addressed by local structure optimization of substrate [16].

It is also observed that the introduction of nonlinearities improves the sensor's performance remarkably. In TABLE III, the residual errors based on each model is analysed. It is found that increasing terms in the equations improves the accuracy and reduces the standard deviation of residual errors significantly, consistent with the theoretical calculation and analysis. The R-squared values of each regression model are determined to check the fitting condition and it is noticed that among these derived models, the model containing more terms have higher R-squared value, which means that higher nonlinearity results in higher fitting goodness thus model's interpretability. Within each model, the equation about the Z axis has lower R-square values and it demonstrates that the regression of Z-axial force components is the most difficult. From TABLE IV, it is noticed that compared to linear calibration, the trained models using the SVM algorithm provides improved performance. In conclusion, utilizing the Gaussian kernel function displays the best force sensing effect. Moreover, when associated with force components along the X and Y axes, its performance is similar to the model derived by quadratic adjusting whilst the performance along the Z axis is optimized remarkably.

## VI. Conclusion

A triaxial force sensor has been successfully developed by using a tapered substrate with an eccentric inner lumen to hold the imaging probe inside and sensing elements outside, as well as maintaining the compatibility with the continuum robot. It enables the triaxial force measurement within a range of 0.5 $N$ along each axis, which has met our design specifications. Different decoupling approaches including linear, nonlinear, regression-based, and learning-based have been developed well to achieve triaxial force sensing. Comparisons of results indicate that the model based on the cross-validation SVM algorithm with the Gaussian kernel function has the best performance, with a satisfactory accuracy of 10.58 $mN$, 14.57 $mN$, 26.32 $mN$ along the X, Y, and Z axes, respectively. To sum up, this design allows a large inner lumen inside to accommodate instruments. It also shows an excellent integration with our developed laser-profiled continuum robot to enable safer optical biopsy.

Future work aims to improve the accuracy along the Z-axis by optimizing the local structure and sensor placement along the circumference. Also, force or stiffness control will be implemented on the laser-profiled continuum robot by using the developed force sensor and the mechanical model [20-24]. Finally, the optical biopsy can be conducted to get high-quality cellular-level images and do the mosaicing in a safer scanning manner.